\documentclass[useAMS,usenatbib,usegraphicx]{mn2e}

\usepackage{times}
\usepackage{longtable}
\usepackage{amsmath}
\usepackage{amssymb}
\usepackage{nopageno}
\usepackage{longtable}

\input{psfig.sty}

\newcommand{\AIPS}{{$\cal AIPS\/$}}

\def\gs{\mathrel{\raise0.35ex\hbox{$\scriptstyle >$}\kern-0.6em
\lower0.40ex\hbox{{$\scriptstyle \sim$}}}}
\def\ls{\mathrel{\raise0.35ex\hbox{$\scriptstyle <$}\kern-0.6em
\lower0.40ex\hbox{{$\scriptstyle \sim$}}}}
\def\m@th{\mathsurround=0pt }
\def\eqalign#1{\null\,\vcenter{\openup1\jot \m@th
 \ialign{\strut\hfil$\displaystyle{##}$&$\displaystyle{{}##}$\hfil
 \crcr#1\crcr}}\,}

\title[The radio spectral index of submm galaxies]
      {Deep multi-frequency radio imaging in the Lockman Hole:
       II.\ The spectral index of submillimetre galaxies}
\author[Ibar et al.]
       {Edo Ibar,$^{\! 1,2}$
	 R.\,J.\ Ivison,$^{\! 1,2}$
	 P.\,N.\ Best,$^{\! 2}$
	 K.\ Coppin,$^{\! 3}$
	 A.\ Pope,$^{\! 4}$ 
	 Ian Smail$^{3}$ and
	 J.\,S.\ Dunlop$^{2}$
         \vspace*{1mm} \\
         $^1$ UK Astronomy Technology Centre, Science and Technology Research Council,
         Royal Observatory, Blackford Hill,
         Edinburgh EH9 3HJ, UK\\
         $^2$ Institute for Astronomy, University of Edinburgh, Blackford Hill,
         Edinburgh EH9 3HJ, UK \\
	 $^3$ Institute for Computational Cosmology, Durham
         University, South Road, Durham DH1 3LE, UK \\
	 $^4$ National Optical Astronomy Observatory, 950 North Cherry
         Avenue, Tucson, AZ 85719, USA \\
	 \vspace{-0.5cm}}

\date{Accepted: 2009 November 9; Received: 2009 October 23; in
  original form: 2009 September 11}

\pagerange{000--000} \pubyear{2009}

\begin{document}

\maketitle

\begin{abstract}
We have employed the Giant Metre-wave Radio Telescope (GMRT) and the
Very Large Array (VLA) to map the Lockman Hole. At 610 and 1,400\,MHz,
we reach noise levels of 15 and 6\,$\mu$Jy\,beam$^{-1}$, respectively,
with well-matched resolutions ($\sim$5\,arcsec). At this depth we
obtained reliable detections for about half of the known submm
galaxies (SMGs) in the field. For radio-identified SMGs, which are
typically at $z\sim2$, we measure a mean radio spectral index of
$\alpha^{\rm 1,400}_{\rm 610} = -0.75\pm0.06$ (where $S_{\nu}\propto
\nu^{\alpha}$) and standard deviation of $0.29$, between approximate
rest-frame frequencies of 1.8 and 4.2\,GHz. The slope of their
continuum emission is indistinguishable from that of local
star-forming galaxies and suggests that extended optically-thin
synchrotron emission dominates the radio output of SMGs. Cooling
effects by synchrotron emission and Inverse Compton (IC) scattering
off the cosmic microwave background (CMB) do not seem to affect their
radio spectral energy distributions (SEDs). For those SMGs judged by
{\em Spitzer} mid-infrared (-IR) colours and spectroscopy to host
obscured active galactic nuclei (AGN), we find a clear deviation from
the rest of the sample -- they typically have steeper radio spectral
indices $\alpha^{\rm 1,400}_{\rm 610} \ls -1.0$.  These findings
suggest these mid-IR-/AGN-selected SMGs may have an intrinsically
different injection mechanism for relativistic particles, or they
might reside in denser environments. This work provides a reliable
spectral template for the estimation of far-IR/radio photometric
redshifts, and will enable accurate statistical $K$-corrections for
the large samples of SMGs expected with SCUBA-2 and {\it Herschel}.
\end{abstract}


\section{Introduction}
\label{intro}

SMGs were discovered in the late 1990s \citep{Smail97} using the
Submillimetre (submm) Common-User Bolometer Array \citep[SCUBA
--][]{Holland99} on the 15-m James Clerk Maxwell Telescope
(JCMT). SCUBA had been designed to exploit the so-called ``negative
$K$-correction'' in the submm waveband, allowing the detection of very
distant dusty galaxies almost unbiased in redshift up to $z\sim\rm
10$. The discovery of these massive, rapidly star-forming galaxies
(star-formation rate -- SFR $\sim10^3\rm M_\odot\, yr^{-1}$), each
potentially capable of creating a massive elliptical galaxy within
1\,Gyr, has provided a powerful motivation for improving models of
galaxy formation and evolution (e.g. \citealt{Swinbank08}). SMGs are
believed by many to be the parent population of present-day elliptical
galaxies and contribute ($S_{850\mu\rm m}>3{\rm mJy}$) for
approximately 20\,per cent of the cosmic IR background (CIRB) at
$850\,\mu$m (\citealt{Eales99}). Understanding the nature of these
galaxies has been extremely challenging, a situation influenced by the
poor resolution of submm images and the resulting confusion at noise
levels required to detect significant numbers of SMGs, and also by
their intrinsically high redshifts (mean $z\approx2.2$ for
radio-detected SMGs -- \citealt{Chapman05b}) and dusty nature which
make optical detections difficult -- radio and IR detections have been
usually employed to address these issues (e.g. \citealt{Ivison07}).

Radio imaging has played a key role in characterising SMGs
\citep[e.g.\ ][]{Ivison02} providing a high-resolution proxy for the
rest-frame far-IR emission via the far-IR/radio correlation
\citep{Helou85, Appleton04, Ibar08}. The radio waveband is a
relatively unexplored part of the SED where current studies have
adopted a canonical power-law form, $S_{\nu}\propto \nu^{\alpha}$,
based on local star-forming galaxies with $\alpha=\rm -0.7$ or $\rm
-0.8$ \citep{Condon92}, and a relatively small dispersion,
$\Delta\alpha\approx0.25$. However, \citet{Hunt05} questioned this
assumption for SMGs, arguing that the adoption of $\alpha=\rm -0.7$ is
not based on any relevant observational evidence. Flatter radio SEDs
-- typical of blue compact dwarf galaxies or some powerful
ultraluminous IR galaxies (ULIRGs, e.g.\ Arp\,220, with $\alpha_{\rm
1.5GHz}^{\rm 8.4GHz}=-0.41$; \citealt{Smith98}) -- were preferred in
\citeauthor{Hunt05}'s photometric redshift analyses of SMGs. Indeed,
\citet{Clemens08} recently measured a mean spectral index of
$\alpha=\rm -0.5$ between 1.4 and 4.8\,GHz for a sample of local
ULIRGs (the local `cousins' of SMGs), although this steepens to
$\alpha=\rm -0.7$ and $-0.8$ between 4.8--8.4 and 8.4--22.5\,GHz,
respectively. The presence of dominant, compact, flat-spectrum,
synchrotron-self-absorbed AGN cores in the SMG population cannot be
ruled out either.

\citet{Kovacs06} employed 350-$\mu$m observations of 15 radio-detected
SMGs with known redshifts to constrain their characteristic
temperatures. In doing so, they claimed to see the first signs of a
deviation from the far-IR/radio correlation exhibited locally, $q_L =
{\rm log}\{L_{\rm FIR}/([{\rm 4.52\,THz}]L_{\rm 1,400MHz})\}\simeq\rm
2.3$ (where $L_{\rm FIR}$ is the total dust luminosity and $L_{\rm
1,400MHz}$ is the rest-frame radio luminosity), the ubiquity of which
has surprised astronomers for decades. \citeauthor{Kovacs06} suggested
that SMGs are over-luminous at radio frequencies with respect to the
far-IR, finding a mean $q_L\approx\rm 2.14\pm 0.07$. Modelling by
\citet{Swinbank08} suggested that this value is not a result of sample
selection but of {\it real} evolution of the far-IR/radio
correlation. \citeauthor{Kovacs06} noted, however, that correcting the
radio spectral indices by $\sim$0.35 would move SMGs back onto the
far-IR/radio correlation (see \S\ref{discu}).

Theoretically, at the redshifts typical of SMGs we expect to observe
steeper radio spectral indices due to the shift of the frequency break
produced by ageing effects (e.g. \citealt{Carilli99}), and/or IC
scattering off the CMB -- going as $(1+z)^4$
(e.g. \citealt{Klamer06}). Indeed, bright radio samples selected on
the basis of their steep ($\alpha\ls\rm -1.0$) spectral indices --
so-called ultra-steep-spectrum (USS) sources -- have been found to
contain a large number of high-redshift galaxies
\citep[e.g.][]{DeBreuck00}. Given their high redshifts, their large
masses and their probable relation to galaxy formation in
proto-clusters, we might expect to find similarities in radio spectral
indices between SMGs and USS sources. We note, however, that there are
major differences between these populations: radio-identified SMGs are
much fainter than typical USS radio galaxies and star formation
(rather than AGN) clearly provides the majority of their power
\citep{Frayer98}. The fraction of SMGs with obvious AGN is relatively
small, although Compton-thick AGN may contribute significantly to the
bolometric emission in some cases. Different approaches have been used
to measure the actual AGN contribution; spectroscopic/photometric
mid-IR diagnostics \citep[e.g.][]{Menendez-Delmestre09}, high
resolution radio imaging (e.g. \citealt{Biggs08}), X-ray
detections/stacking (e.g. \citealt{Alexander03}), and deviations from
the FIR/radio correlation (e.g. \citealt{Kovacs06}).

We thus have strong motives for exploring the radio spectral index of
SMGs: to provide an independent description of their radio SEDs, to
enable more accurate radio $K$-corrections, to better test whether the
far-IR/radio correlation evolves, and to find possible new AGN
diagnostics. Here, we report deep, dual-frequency, matched-resolution
radio observations in the Lockman Hole that were designed to address
these issues.

\section{Multi-wavelength observations in the Lockman Hole}

\subsection{Parent catalogue of submm galaxies}
\label{submm_obs}

As one of the SCUBA Half Degree Extragalactic Survey (SHADES) fields,
485\,arcmin$^2$ of the Lockman Hole was observed with SCUBA on JCMT at
850\,$\mu$m, producing images with a {\sc fwhm} beamsize of
14.8\,arcsec. In this work, we use the catalogue provided by
\citet{Coppin06}, comprising 57 sources detected with a
signal-to-noise ratio, $\rm SNR \geq 3.5$ in more than one independent
reduction of the data. The sample includes only sources with a low
probability ($\le$5 per cent) that the true, deboosted flux density is
lower than zero (i.e.\ that the source is entirely spurious).

The Lockman Hole was also observed using the Astronomical Thermal
Emission Camera \citep[AzTEC --][]{Wilson08} on JCMT at 1,100\,$\mu$m
\citep{Austermann09}. This survey covers a large area:
1,115\,arcmin$^2$, to an r.m.s.\ of $\sim$0.9--1.3\,mJy\,beam$^{-1}$
($\sim$18\,arcsec {\sc fwhm}). In this work, we use the
\citeauthor{Austermann09} sample of 51 sources with $\rm SNR \geq 3.5$
and $P(S_{\rm 1,100\mu m}\leq0\,\rm mJy)<0.05$, i.e.\ the same
definition used by \citeauthor{Coppin06}.

The MAx-Planck Millimeter BOlometer array
\citep[MAMBO-117][]{Kreysa99} on the Instituto de Radioastronomia
Milim\'etrica (IRAM) 30-m telescope has also been used to image the
Lockman Hole. \citet{Greve04} acquired data at 1,200\,$\mu$m, covering
197\,arcmin$^2$ down to an r.m.s.\ of 0.6\,mJy\,beam$^{-1}$
(11\,arcsec, {\sc fwhm}). We use the \citeauthor{Greve04} catalogue
which contains 23 sources with $\rm SNR \geq 3.5$.

The combination of AzTEC, SCUBA and MAMBO imaging makes the Lockman
Hole the largest area of sky mapped at 850, 1,100 and 1,200\,$\mu$m.
Combining the three samples results in an overall sample of 111 SMGs
(20 of which were detected in more than one sample) which we have
adopted for cross-matching with our radio catalogues. We opted not to
use the Bolocam catalogue by \citet{Laurent05} because of the
confusion resulting from the large beam ($\sim$31\,arcsec {\sc fwhm}).

\subsubsection{AGN in the sample}
\label{agnes}

Spectroscopic analyses using {\it Spitzer}'s IR Spectrograph
\citep[IRS --][]{Houck04} to identify the mid-IR power source in SMGs
\citep[e.g.][]{Pope08, Menendez-Delmestre09} have confirmed earlier
indications that they are usually powered by extreme star formation
rather than by AGN activity. Indeed, based on composite median
template of a sample of 24 radio-identified SMGs,
\citeauthor{Menendez-Delmestre09} find a typical AGN contribution of
$<32\,$per cent to the bolometric mid-IR emission.

However, some of the AzTEC and SCUBA sources in the SHADES fields were
selected for IRS observations using colour criteria designed to
preferentially select highly obscured AGN (\citeauthor{Coppin06}, in
preparation). A sample of six such sources was targeted in the Lockman
Hole, from which four have a dominant ($>$50-per-cent) AGN
contribution to their mid-IR spectra. We use this sub-sample to find
possible deviations in the behaviour of radio spectral indices (see
\S\ref{a_smg}).

\subsection{GMRT and VLA imaging}
\label{radio_maps}

We have obtained deep radio observations in the Lockman Hole using the
GMRT\footnote{GMRT is run by the National Centre for Radio
Astrophysics of the Tata Institute of Fundamental Research.} and the
National Radio Astronomy Observatory's (NRAO) VLA\footnote{NRAO is
operated by Associated Universities Inc., under a cooperative
agreement with the National Science Fundation.}, operated at 610 and
1,400\,MHz, respectively. The two image mosaics have r.m.s.\ noise
levels in the central regions of 15 and 6\,$\mu$Jy\,beam$^{-1}$ and
synthesised beams ({\sc fwhm}) of $7.1\times6.5$ and
$4.3\times4.2$\,arcsec$^2$, respectively (see a full description in
\citealt{Ibar09}).

In this work, we investigate the radio spectral indices of the SMG
sample described in \S\ref{submm_obs}, exploiting our Lockman Hole
radio data down to a peak-to-noise ratio, $\rm PNR \geq 3$ -- where
peak and noise refer to the maximum value of a 2-D Gaussian fit and
the local r.m.s., respectively. Given the resolution of our radio
imaging (4--7\,arcsec, {\sc fwhm}) and the typical kpc scale of SMGs
\citep[e.g.][]{Biggs08}, we have considered the SMG counterparts to be
unresolved. We have extracted flux densities with the source size in
the {\sc sad} Gaussian fitting routine within the Astronomical Image
Processing System\footnote{see http://www.aips.nrao.edu/cook.html for
a description of tasks.} (\AIPS) set to the size of the synthesised
beam ({\sc dowidth} = $-$1), which results in a less uncertain
integrated flux density estimate because of the lower number of free
parameters in the Gaussian fit. Monte Carlo simulations based on fake
Gaussians injected in a residual map (with no $>$\,5-$\sigma$ peaks)
using {\sc immod} and extracted using {\sc sad} show a clear
improvement ($\sim$1.2$\times$ less scatter in the out:in flux density
ratios at $\rm PNR < 5$) by fixing the synthesised beam in this
way. Fewer outliers are seen as well.

There is an important caveat concerning the use of these faint,
low-PNR sources: a very large fraction of the sources detected at
faint flux densities are due to the background noise. To determine the
prevalence of spurious sources in the samples, we inspected the number
of $\geq$\,3-$\sigma$ sources extracted from the inverted residual
map. The cumulative number of negative sources, $N_{\rm neg}($$>$$S)$,
increased rapidly toward faint flux densities, overwhelming the number
counts of the 'positive' sources at $S_{\rm 1,400MHz}\ls40\,\mu$Jy and
$S_{\rm 610MHz}\ls120\,\mu$Jy. This indicates that an extrapolation of
the number counts to $S_{\rm 1,400MHz}\ls 100\,\mu$Jy ($S_{\rm
610MHz}\ls 300\,\mu$Jy), based on sources with higher flux densities,
will thus underestimate the number of extracted sources in the
field. This is relevant as we need to know the probability of random
associations between SMGs and faint radio sources.

\subsection{Radio identifications}
\label{rad_id}

Associating SMGs with radio counterparts has usually been accomplished
using probabilistic arguments (e.g. \citealt{Ivison07}). We have
chosen to cross-match the parent SMG catalogue with our radio maps
(\S\ref{radio_maps}) using a $P<0.05$ selection criterion, meaning
that the probability of a spurious association is less than 5 per cent
(\citealt{Downes86}). This probability has two dependencies -- the
radius within which we search for radio counterparts and the
cumulative radio source number count distribution per surface unit,
$N($$>$$S)$. The search radius was chosen to be 3$\times$ the
uncertainty in the submm position (in Right Ascension or Declination),
as estimated using Eq.\,B22 of \citet{Ivison07}, i.e.\ dependent on
the submm SNR, with 12.0, 14.4 and 8.8\,arcsec for $\rm SNR=4$ SCUBA,
AzTEC and MAMBO detections; we expect 1 per cent of the true
counterparts to lie outside our search area. For the number count
dependency, we used all $\geq$\,3-$\sigma$ radio detections to find
the true probability of random association. Nevertheless, if the
source was detected at both radio wavelengths (unlikely to be noise)
we used a negative-subtracted cumulative radio source number count
distribution instead, $N_{\rm sub}($$>$$S)=N($$>$$S)-N_{\rm
neg}($$>$$S)$, in order to give a better proxy for the number of real
sources. We note, however, that the power of this $P$-identification
procedure is lessened for very large submm beams and/or low SNRs
because the purely statistical approach is unable to discriminate
against unrelated radio-bright sources (preferentially radio-loud AGN)
at large separations.

We adopted a conservative approach, eliminating from the analysis all
SMGs with more than two radio sources that could affect the
measurement of the radio spectral index, thus biasing our sample
against close mergers. A final inspection by eye was used to eliminate
sources with confused associations. The final sample comprised 44 SMGs
(from a total of 111) with single and reliable radio counterparts in
at least one radio waveband (41 and 37 detections by the VLA and GMRT,
respectively).

\begin{figure*}
  \begin{center}
  \includegraphics[scale=0.53]{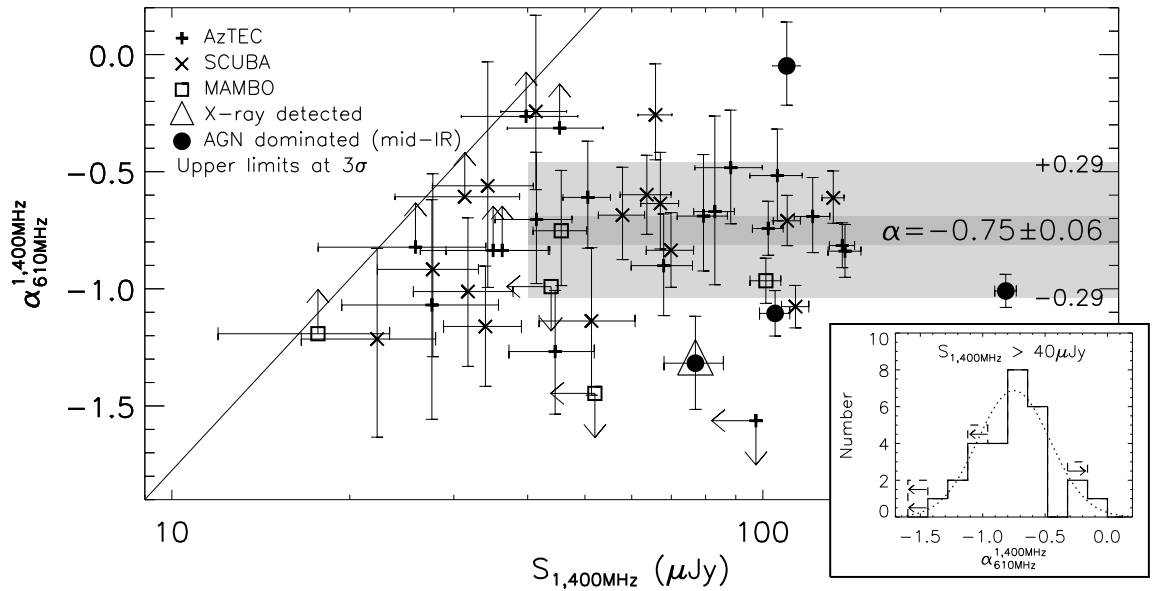}
  \caption{The radio spectral index, between 610 and 1,400\,MHz, of a
    sample of 44 SMGs selected from the Lockman Hole as a function of
    observed radio flux density at 1,400\,MHz. Plus, cross and square
    symbols represent unique AzTEC, SCUBA and MAMBO sources (see
    \S\,\ref{rad_id}). Error bars are estimated from the uncertainty
    in flux density at each frequency. Upper and lower limits are
    3\,$\sigma$ thresholds. Large filled circles are SMGs dominated by
    AGN activity as judged by {\em Spitzer} mid-IR colours and
    spectroscopy (\citeauthor{Coppin06}, in preparation). A cross
    shows the source is detected in X-rays. The solid line denotes the
    threshold in radio spectral index based on a 3-$\sigma$ source
    with $S_{\rm 610MHz}=45\,\mu$Jy. The mean and standard deviation
    (using only $S_{\rm 1,400MHz}>40\,\mu$Jy sources) are shown in
    dark and light grey, respectively. The box at the bottom-right
    corner shows the histogram of radio spectral indexes for those
    sources with $S_{\rm 1,400MHz}>40\,\mu$Jy; solid, dashed and
    dotted lines are detections, upper limits and the normal
    distribution fit, respectively.}
  \label{flux_alpha}
\end{center}
\end{figure*}

\section{The radio spectral index of SMGs}
\label{a_smg}

\subsection{Observations}

In Fig.~\ref{flux_alpha}, we show the radio spectral indices of the 44
radio-selected SMGs as a function of flux density at 1,400\,MHz. The
vast majority of these sources were detected at low flux densities,
$S_{\rm 1,400MHz}\ls 150\,\mu$Jy, close to the detection limits of
even the deepest radio imaging.

The study of the radio spectral index has a clear bias introduced by
the detection threshold in each map. The solid black line in
Fig.~\ref{flux_alpha} shows the restriction for detecting
flat-spectrum sources at faint 1,400-MHz radio fluxes, based on a
minimum $3$-$\sigma$ source at $S_{\rm 610MHz}\approx 45\,\mu$Jy in
the deepest region of the 610\,MHz image (note that noise varies
within a factor of $\ls$\,2 with position in the maps). In order to
reduce the effect of this bias, we have restricted the statistical
analysis to those sources with $S_{\rm 1,400MHz}>40\,\mu$Jy only (see
shaded area in Fig.~\ref{flux_alpha}). From this sub-sample, 28 were
detected at both radio wavelengths, while one and three were detected
only by the VLA and GMRT, respectively. We find -- for those SMGs
detected at both radio frequencies -- a bootstrapped median radio
spectral index of $-0.71\pm0.04$, a mean of $-0.75\pm0.06$ and
standard deviation of $0.29$. In the data, there is no clear
correlation between $\alpha^{\rm 1,400}_{\rm 610}$, radio and submm
flux densities, suggesting the radio spectral indices of the
radio-identified SMGs shown in this work may provide a good template
for SMGs below the current submm confusion limit.

By looking at the sample of four AGNs selected via mid-IR
spectroscopic analysis (\S\ref{agnes}; \citeauthor{Coppin06}, in
preparation), we find all of them have a reliable
$\alpha_{610}^{1,400}$ measurement (\S\ref{rad_id}). Although a small
sample, restricted to bright mid-IR fluxes, their radio properties
clearly deviate from those of the other SMGs (filled black circles in
Fig.~\ref{flux_alpha}). One is a flat-spectrum source,
$\alpha_{610}^{1,400}=-0.05$, while the other three present steep
radio spectra, $\alpha_{610}^{1,400}\ls -1.0$.

The different nature of these sources is also evidenced by the
detection of the steepest radio source (mid-IR AGN-dominated) in
X-rays (\citealt{Brunner08}) showing clear evidence of a highly
obscured AGN -- in agreement with \citeauthor{Coppin06}'s criteria --
we note this is the only SMG with a clear X-ray detection. Flat radio
spectra are also related to blazar or powerful compact AGNs and are
usually interpreted as the synchrotron self-absorbed base of a
``young'' expanding jet that fuels larger scale radio emission lobes.

\section{Discussion}
\label{discu}

The observed radio spectral indices for SMGs, at approximately
rest-frame 1.8 -- 4.2\,GHz for $z\sim2$, are similar to either those
found in local star-forming galaxies (\citealt{Condon92}) and to the
bulk of the sub-mJy radio sources found by \citet{Ibar09}. The flux
density range at which the vast majority of the SMGs are found,
$S_{\rm 1,400MHz}<150\,\mu$Jy, suggests they may compose a significant
fraction of the radio population responsible for the flattening of the
Euclidean radio number counts.

Assuming the physics of the synchrotron emission does not evolve as a
function of redshift, these results suggest that non-thermal optically
thin synchrotron emission dominates in the majority of SMGs. This
suggests magnetic fields are large enough to retain cosmic rays (see
\citealt{Thompson06}) and contradicts the idea that flatter radio SEDs
-- such as those thermally dominated galaxies \citep{Hunt05} or in
synchrotron self-absorbed sources -- are more appropriate SEDs for
these massive star-forming galaxies. Indeed, the lack of flat spectrum
sources suggests that their emission is mostly dominated by extended
rather than compact radio structures, in agreement with previous SMG
high-resolution radio observations (\citealt{Chapmanmerlin,Biggs08})
and the low mid-IR extinctions found by \citet{Menendez-Delmestre09}.

The lack of evidence for spectral steepening with respect to local
star-forming galaxies suggests that $K$-correction effects in curved
SEDs and IC scattering enhanced by a denser CMB radiation field at
high redshift (\citealt{Klamer06,Miley08}) do not have a strong
influence on the observed radio spectral index of these high-redshift
galaxies. The possibility of intrinsic redshift evolution to flatter
radio spectra, where curved SEDs (due to synchrotron and IC cooling)
steepen them back to the observed values, cannot be ruled out.

The steeper radio spectral indices found for the mid-IR AGN-dominated
sources are intriguing. Recent studies of USS samples (also at high
redshift and mostly AGN) using multi-frequency radio observations have
found no spectral curvature at higher frequencies in these
samples. The power-law behaviour suggests steep radio spectra in
high-redshift AGN are intrinsic, or a product of environment
\citep{Bryant09}. The former relates to a different mechanism for the
injection of the initial electron energy distribution \citep[e.g.\
][]{Jaffe73, Carilli99} which determine the reservoir of high-energy
particles -- we expect to observe steeper radio spectra if there is no
continuous injection or re-acceleration of relativistic particles
across the galaxy. On the other hand, if the source is embedded in a
dense medium, some of the energy of the particles would be released
into the environment and stronger magnetic fields may increase the
cooling rate (\citealt{Thompson06}), i.e.\ the radio SED would become
steeper.

Based on the fact that we do not observe steeper radio spectral
indices in the bulk of the SMGs, these findings may suggest that the
synchrotron emission in AGN is fundamentally different to the emission
seen in massive star-forming galaxies. If we assume the $\alpha$ vs
redshift correlation found in brighter radio samples
(\citealt{DeBreuck00}), it is possible that the synchrotron radiation
emitted by nuclear activity evolves differently as a function of
cosmic time than that emitted due to star formation.  Nevertheless,
there are major differences between USS and SMGs as pointed out in
\S\ref{intro}.

We have checked the possibility for an overestimation of the radio
luminosities in the sample analysed by \citet{Kovacs06}. From the 9
SMGs in the Lockman Hole (Table~1 from their paper), 5 of them have
got a reliable estimate of radio spectral index, $\alpha_{\rm
610}^{\rm 1,400}=-0.68\pm0.26$ (bootstrapped median). Based on these
detections, it seems that their sample is not particularly affected by
peculiar radio $K$-corrections that may explain the observed deviation
from the local far-IR/radio correlation.

A future analysis -- using radio observations from the LOw Frequency
ARray (LOFAR) and Expanded VLA to cover a wide range of frequency, for
samples with well-constrained redshifts -- will allow us to give a
much better description of the synchrotron emission processes in
massive star-forming galaxies and radio faint AGN. An independent
assessment can also be obtained using high-resolution radio
observations in order to determine the physical extent, morphology,
luminosity and brightness temperature of their emitting regions.

\section{Conclusions}

We have analysed the radio spectral indices, $\alpha_{\rm 610}^{\rm
1,400}$, based on well-matched observations with GMRT at 610\,MHz and
the VLA at 1,400\,MHz, of a sample of SMGs selected in the Lockman
Hole using the AzTEC, SCUBA, and MAMBO cameras.

We have created a sample of 44 SMGs (from a total of 111 unique SMGs)
which have secure radio identifications and reliable estimates of
radio spectral index. We report how deep GMRT observations have
resulted in a detection rate similar to that usually accomplished with
the VLA. We find a mean value of $\alpha_{\rm 610}^{\rm
1,400}=-0.75\pm0.06$ (see Fig.~\ref{flux_alpha}), suggesting the
majority of SMGs are dominated by optically thin synchrotron emission
from extended radio structures; star-forming regions and/or AGN
lobes. The distribution of spectral indices for SMGs is
indistinguishable from that of local star-forming galaxies. This
supports the idea that SMGs are dominated by processes typical of
dusty star-forming galaxies rather than those seen in AGN. If we
assume that there is no intrinsic redshift evolution for the
synchrotron emission mechanism, we find that ageing effects and IC
scattering off the CMB does not seem to largely affect -- steepen --
the bulk of the observed radio spectral indices. We do not observe any
dependency of the radio spectral index on the submm/radio flux density
ratio, or on submm or radio flux densities, which suggests the radio
spectral properties found in this work should also be representative
of the SMGs below the current confusion limits.

We have found that the radio spectral indices of SMGs with an
AGN-dominated mid-IR spectrum deviate from the bulk of the
sample. They typically have either steeper radio spectral indices,
$\alpha^{\rm 1,400}_{\rm 610}\ls -1.0$ -- similar to USS sources found
in bright radio samples -- or flat spectra. These findings suggest a
different mechanism responsible for the synchrotron emission in these
sources: either related to the injection mechanism of relativistic
plasma, or a product of the environment. Their deviation to steeper
spectral indices could be an evidence for a different cosmic radio
evolution with respect to massive star-forming galaxies. Besides of
the small sample of sources, radio spectral index thus represents a
useful selection criterion for identifying AGN in high-redshift SMGs.

Looking to the future, these results can be used for statistical
analyses of the large SMG samples anticipated from SCUBA-2 and from
{\em Herschel}. Our parameterisation of the radio SED of SMGs will
allow reliable $K$-corrections for flux densities in the $\sim$1-GHz
regime, and will alleviate degeneracies faced by long-wavelength
photometric redshift techniques.

\section*{Acknowledgements}

This paper was supported by a Gemini research studentship. We thank
the staff of the GMRT for making these observations possible. EI, RJI,
PNB, KEKC, IRS and JSD acknowledge support from UK Science and
Technology Research Council.

\setlength{\labelwidth}{0pt} 

\bibliographystyle{mn2e}
\bibliography{ibar}

\begin{thebibliography}{}

\bibitem[\protect\citeauthoryear{{Alexander} et~al.}{{Alexander}
  et~al.}{2003}]{Alexander03}
{Alexander} D.~M. et~al., 2003, \aj, 125, 383

\bibitem[\protect\citeauthoryear{{Appleton} et~al.}{{Appleton}
  et~al.}{2004}]{Appleton04}
{Appleton} P.~N. et~al., 2004, \apjs, 154, 147

\bibitem[\protect\citeauthoryear{{Austermann} et~al.}{{Austermann}
  et~al.}{2009}]{Austermann09}
{Austermann} J.~E. et~al., 2009, ArXiv e-prints

\bibitem[\protect\citeauthoryear{{Biggs} \& {Ivison}}{{Biggs} \&
  {Ivison}}{2008}]{Biggs08}
{Biggs} A.~D.,  {Ivison} R.~J., 2008, \mnras, 385, 893

\bibitem[\protect\citeauthoryear{{Brunner} et~al.}{{Brunner}
  et~al.}{2008}]{Brunner08}
{Brunner} H., {Cappelluti} N., {Hasinger} G., {Barcons} X., {Fabian} A.~C.,
  {Mainieri} V.,  {Szokoly} G., 2008, \aap, 479, 283

\bibitem[\protect\citeauthoryear{{Bryant} et~al.}{{Bryant}
  et~al.}{2009}]{Bryant09}
{Bryant} J.~J., {Johnston} H.~M., {Broderick} J.~W., {Hunstead} R.~W., {De
  Breuck} C.,  {Gaensler} B.~M., 2009, \mnras, 395, 1099

\bibitem[\protect\citeauthoryear{{Carilli} \& {Yun}}{{Carilli} \&
  {Yun}}{1999}]{Carilli99}
{Carilli} C.~L.,  {Yun} M.~S., 1999, \apjl, 513, L13

\bibitem[\protect\citeauthoryear{{Chapman} et~al.}{{Chapman}
  et~al.}{2005}]{Chapman05b}
{Chapman} S.~C., {Blain} A.~W., {Smail} I.,  {Ivison} R.~J., 2005, \apj, 622,
  772

\bibitem[\protect\citeauthoryear{{Chapman} et~al.}{{Chapman}
  et~al.}{2004}]{Chapmanmerlin}
{Chapman} S.~C., {Smail} I., {Windhorst} R., {Muxlow} T.,  {Ivison} R.~J.,
  2004, \apj, 611, 732

\bibitem[\protect\citeauthoryear{{Clemens} et~al.}{{Clemens}
  et~al.}{2008}]{Clemens08}
{Clemens} M.~S., {Vega} O., {Bressan} A., {Granato} G.~L., {Silva} L.,
  {Panuzzo} P., 2008, \aap, 477, 95

\bibitem[\protect\citeauthoryear{{Condon}}{{Condon}}{1992}]{Condon92}
{Condon} J.~J., 1992, \araa, 30, 575

\bibitem[\protect\citeauthoryear{{Coppin} et~al.}{{Coppin}
  et~al.}{2006}]{Coppin06}
{Coppin} K. et~al., 2006, \mnras, 372, 1621

\bibitem[\protect\citeauthoryear{{De Breuck} et~al.}{{De Breuck}
  et~al.}{2000}]{DeBreuck00}
{De Breuck} C., {van Breugel} W., {R{\"o}ttgering} H.~J.~A.,  {Miley} G., 2000,
  \aaps, 143, 303

\bibitem[\protect\citeauthoryear{{Downes} et~al.}{{Downes}
  et~al.}{1986}]{Downes86}
{Downes} A.~J.~B., {Peacock} J.~A., {Savage} A.,  {Carrie} D.~R., 1986, \mnras,
  218, 31

\bibitem[\protect\citeauthoryear{{Eales} et~al.}{{Eales}
  et~al.}{1999}]{Eales99}
{Eales} S., {Lilly} S., {Gear} W., {Dunne} L., {Bond} J.~R., {Hammer} F., {Le
  F{\`e}vre} O.,  {Crampton} D., 1999, \apj, 515, 518

\bibitem[\protect\citeauthoryear{{Frayer} et~al.}{{Frayer}
  et~al.}{1998}]{Frayer98}
{Frayer} D.~T., {Ivison} R.~J., {Scoville} N.~Z., {Yun} M., {Evans} A.~S.,
  {Smail} I., {Blain} A.~W.,  {Kneib} J.-P., 1998, \apjl, 506, L7

\bibitem[\protect\citeauthoryear{{Greve} et~al.}{{Greve}
  et~al.}{2004}]{Greve04}
{Greve} T.~R., {Ivison} R.~J., {Bertoldi} F., {Stevens} J.~A., {Dunlop} J.~S.,
  {Lutz} D.,  {Carilli} C.~L., 2004, \mnras, 354, 779

\bibitem[\protect\citeauthoryear{{Helou}, {Soifer} \& {Rowan-Robinson}}{{Helou}
  et~al.}{1985}]{Helou85}
{Helou} G., {Soifer} B.~T.,  {Rowan-Robinson} M., 1985, \apjl, 298, L7

\bibitem[\protect\citeauthoryear{{Holland} et~al.}{{Holland}
  et~al.}{1999}]{Holland99}
{Holland} W.~S. et~al., 1999, \mnras, 303, 659

\bibitem[\protect\citeauthoryear{{Houck} et~al.}{{Houck}
  et~al.}{2004}]{Houck04}
{Houck} J.~R. et~al., 2004, in Presented at the Society of Photo-Optical
  Instrumentation Engineers (SPIE) Conference, Vol. 5487, {Mather} J.~C., ed,
  Society of Photo-Optical Instrumentation Engineers (SPIE) Conference Series,
  p.~62

\bibitem[\protect\citeauthoryear{{Hunt} \& {Maiolino}}{{Hunt} \&
  {Maiolino}}{2005}]{Hunt05}
{Hunt} L.~K.,  {Maiolino} R., 2005, \apjl, 626, L15

\bibitem[\protect\citeauthoryear{{Ibar} et~al.}{{Ibar} et~al.}{2008}]{Ibar08}
{Ibar} E. et~al., 2008, \mnras, 386, 953

\bibitem[\protect\citeauthoryear{{Ibar} et~al.}{{Ibar} et~al.}{2009}]{Ibar09}
{Ibar} E., {Ivison} R.~J., {Biggs} A.~D., {Lal} D.~V., {Best} P.~N.,  {Green}
  D.~A., 2009, \mnras, 397, 281

\bibitem[\protect\citeauthoryear{{Ivison} et~al.}{{Ivison}
  et~al.}{2007}]{Ivison07}
{Ivison} R.~J. et~al., 2007, \mnras, 380, 199

\bibitem[\protect\citeauthoryear{{Ivison} et~al.}{{Ivison}
  et~al.}{2002}]{Ivison02}
{Ivison} R.~J. et~al., 2002, \mnras, 337, 1

\bibitem[\protect\citeauthoryear{{Jaffe} \& {Perola}}{{Jaffe} \&
  {Perola}}{1973}]{Jaffe73}
{Jaffe} W.~J.,  {Perola} G.~C., 1973, \aap, 26, 423

\bibitem[\protect\citeauthoryear{{Klamer} et~al.}{{Klamer}
  et~al.}{2006}]{Klamer06}
{Klamer} I.~J., {Ekers} R.~D., {Bryant} J.~J., {Hunstead} R.~W., {Sadler}
  E.~M.,  {De Breuck} C., 2006, \mnras, 371, 852

\bibitem[\protect\citeauthoryear{{Kov{\'a}cs} et~al.}{{Kov{\'a}cs}
  et~al.}{2006}]{Kovacs06}
{Kov{\'a}cs} A., {Chapman} S.~C., {Dowell} C.~D., {Blain} A.~W., {Ivison}
  R.~J., {Smail} I.,  {Phillips} T.~G., 2006, \apj, 650, 592

\bibitem[\protect\citeauthoryear{{Kreysa} et~al.}{{Kreysa}
  et~al.}{1999}]{Kreysa99}
{Kreysa} E. et~al., 1999, Infrared Physics and Technology, 40, 191

\bibitem[\protect\citeauthoryear{{Laurent} et~al.}{{Laurent}
  et~al.}{2005}]{Laurent05}
{Laurent} G.~T. et~al., 2005, \apj, 623, 742

\bibitem[\protect\citeauthoryear{{Men{\'e}ndez-Delmestre}
  et~al.}{{Men{\'e}ndez-Delmestre} et~al.}{2009}]{Menendez-Delmestre09}
{Men{\'e}ndez-Delmestre} K. et~al., 2009, \apj, 699, 667

\bibitem[\protect\citeauthoryear{{Miley} \& {De Breuck}}{{Miley} \& {De
  Breuck}}{2008}]{Miley08}
{Miley} G.,  {De Breuck} C., 2008, \aapr, 15, 67

\bibitem[\protect\citeauthoryear{{Pope} et~al.}{{Pope} et~al.}{2008}]{Pope08}
{Pope} A. et~al., 2008, \apj, 689, 127

\bibitem[\protect\citeauthoryear{{Smail}, {Ivison} \& {Blain}}{{Smail}
  et~al.}{1997}]{Smail97}
{Smail} I., {Ivison} R.~J.,  {Blain} A.~W., 1997, \apjl, 490, L5

\bibitem[\protect\citeauthoryear{{Smith}, {Lonsdale} \& {Lonsdale}}{{Smith}
  et~al.}{1998}]{Smith98}
{Smith} H.~E., {Lonsdale} C.~J.,  {Lonsdale} C.~J., 1998, \apj, 492, 137

\bibitem[\protect\citeauthoryear{{Swinbank} et~al.}{{Swinbank}
  et~al.}{2008}]{Swinbank08}
{Swinbank} A.~M. et~al., 2008, \mnras, 391, 420

\bibitem[\protect\citeauthoryear{{Thompson} et~al.}{{Thompson}
  et~al.}{2006}]{Thompson06}
{Thompson} T.~A., {Quataert} E., {Waxman} E., {Murray} N.,  {Martin} C.~L.,
  2006, \apj, 645, 186

\bibitem[\protect\citeauthoryear{{Wilson} et~al.}{{Wilson}
  et~al.}{2008}]{Wilson08}
{Wilson} G.~W. et~al., 2008, \mnras, 386, 807

\end{thebibliography}

\end{document}